\documentclass[aps,pra,reprint,superscriptaddress]{revtex4-2}
\usepackage[normalem]{ulem}               
\usepackage{siunitx}
\usepackage{graphicx}
\usepackage{dcolumn}
\usepackage{bm}
\usepackage[utf8]{inputenc}
\usepackage[T1]{fontenc}
\usepackage{etoolbox}
\usepackage[caption=false]{subfig}
\usepackage{newtxtext,newtxmath}
\RequirePackage{fix-cm}
\DeclareSIUnit\bar{bar}
\DeclareSIUnit{\belmilliwatt}{Bm}
\DeclareSIUnit{\dBm}{\deci\belmilliwatt}
\DeclareSIUnit\bar{bar}
\usepackage{hyperref}
\begin{document}

\title{Extracting higher-order nonlinearities in nanomechanical resonators using the backbone relation} 

\author{Maria Kallergi}
\affiliation{Department of Electrical Engineering, School of Computation, Information and Technology,
Technical University of Munich, 85748 Garching, Germany}

\author{Daniel K.~J. Bone{\ss}}
\affiliation{Department of Physics, University of Konstanz, 78457 Konstanz, Germany}

\author{Maximilian Seitner}
\affiliation{Department of Physics, University of Konstanz, 78457 Konstanz, Germany}

\author{Wolfgang Belzig}
\affiliation{Department of Physics, University of Konstanz, 78457 Konstanz, Germany}

\author{Eva M. Weig}
\affiliation{Department of Electrical Engineering, School of Computation, Information and Technology,
Technical University of Munich, 85748 Garching, Germany}
\affiliation{Munich Center for Quantum Science and Technology (MCQST), 80799 Munich, Germany}
\affiliation{TUM Center for Quantum Engineering (ZQE), 85748 Garching, Germany}

\date{\today}

\begin{abstract} 
 Nanomechanical resonators are a powerful platform for studying nonlinear dynamics with high sensitivity and precision. We explore the nonlinear response of a high-Q nanomechanical string resonator in and beyond the Duffing regime and introduce a robust framework for accurately extracting its conservative nonlinearities. The method is based on the backbone curve obtained from ringdown measurements, making it inherently resilient to small frequency fluctuations while explicitly accounting for both symmetry-breaking and non-symmetry-breaking nonlinearities. To validate the approach, we perform complementary ringdown and frequency-response measurements on the nanostring resonator and benchmark the backbone-based extraction against established frequency-response techniques. The comparison confirms the accuracy of the proposed framework and demonstrates its advantages over conventional methods for nonlinear characterization.
\end{abstract}


\maketitle 

The field of nanoelectromechanical systems (NEMS) has established itself as a key platform for a wide range of applications~\cite{Bachtold2022}, while advances in material and geometry engineering have enabled extremely high quality factors (Q)~\cite{Imboden2014, Sementilli2021, Engelsen2024}. The realization of weakly damped, high‑Q nanomechanical resonators allows large oscillation amplitudes even under small drive power, bringing devices into the nonlinear regime at moderate driving forces and thereby opening access to a rich landscape of nonlinear and out‑of‑equilibrium phenomena~\cite{lifshitz2008nonlinear,Bachtold2022}. 

Representative examples include noise‑enabled precision measurements near the Duffing bifurcation \cite{Aldridge2005}, nonlinear switching dynamics at the onset of bistability \cite{Unterreithmeier2010}, and noise‑induced switching in bistable and metastable states \cite{Chan2007,Chan2008paths, Venstra2013, Chowdhury2017}. The Duffing nonlinearity also enables spectral squeezing of thermal fluctuations in driven nanomechanical modes \cite{huber2020spectral, ochs2021amplification, Almog2007, Suh2010, Mahboob2010, Rugar1991, Yang2021}, as well as amplification and signal enhancement via careful control of bifurcation topology \cite{Karabalin2011}. Higher-order nonlinearities have been identified and exploited for frequency or amplitude stabilization~\cite{Kacem2009, polunin2016characterization,Huang2019,Miller2021, Kaisar2022}. 

Beyond single-mode nonlinearities, nonlinear intermodal couplings also impact the resonator dynamics. Both off-resonant nonlinear mode coupling~\cite{Westra2010,Matheny2013} and resonant nonlinear mode coupling at internal resonances~\cite{antonio2012frequency,Eichler2012} have been experimentally demonstrated, and subsequently exploited to achieve functionalities such as efficient energy transfer~\cite{Chen2017,Yan2022}, phase locking~\cite{Wang2022}, and the generation of nanomechanical frequency combs~\cite{Czaplewski2018}.

Besides the conservative nonlinearities discussed so far, dissipative nonlinearities are also found under large amplitude vibrations, giving rise to nonlinear damping~\cite{Eichler2011, Zaitsev2011, polunin2016characterization}.
This variety of nonlinear coefficients and phenomena underscores the need for a reliable method to quantify the effective nonlinear parameters in a systematic and experimentally robust way.

Conventionally, the dynamics of an individual resonator mode are modeled by a single coordinate governed by an anharmonic potential. This potential can be expanded in the form 

\begin{equation}
U(q) = \frac{M}{2} \omega^2_{0} q^2 + \frac{M}{3} \gamma_{3} q^3+ \frac{M}{4} \gamma_{4} q^4 + \frac{M}{5} \gamma_{5} q^5 + \frac{M}{6} \gamma_{6} q^6 + .... \label{eq:fullpot}
\end{equation}

with single-mode displacement $q$, effective mass $M$, eigenfrequency $\omega_0$, and nonlinear expansion coefficients $\gamma_i$ ($i \geq 3$).

However, generally, a resonator hosts many different modes that are coupled to one another. Such systems are then better described using the action-angle formalism~\cite{Landau1976,Goldstein2002}, where the mode displacement and the corresponding momentum are replaced by an action and a phase variable. By definition, the Hamiltonian is independent of the phase variables. In the regime where only one mode is excited, the action coordinates of the other modes are small. The Hamiltonian can then be approximated as $H(I) \approx \int \omega_\mathrm{I}(I)\, dI$ with the nonlinear eigenfrequency given by $\omega_\mathrm{I}(I) = \partial H/\partial I$ and $I$ the action of the excited mode. The frequency $\omega_\mathrm{I}(I)$ is then usually expanded in powers of $I$. The corresponding expansion coefficients already include all the nonlinearities that may arise, e.g., from the anharmonic potential of the mode or intermodal couplings.

In most cases, capturing all the other modes and determining their coupling to the driven mode is not feasible. The conventional approach is to start with a microscopic model that only assumes a nonlinear potential for the driven mode. The expansion of such a potential is described in Eq.~(\ref{eq:fullpot}). Here, the first, quadratic term in this Taylor approximation describes the harmonic response, while the higher-order terms account for conservative nonlinearities in the system. For small vibration amplitudes, the harmonic approximation is sufficient to describe the resonator dynamics. However, as the amplitude increases, higher-order contributions become progressively more significant and must be taken into account.

In many resonators, the underlying structure exhibits spatial symmetry. Under such conditions, symmetry-breaking odd-order nonlinear coefficients $\gamma_{2i-1}$ vanish, and the potential is fully described by the even-order terms $\gamma_{2i}$ ($i \geq 2$)~\cite{lifshitz2008nonlinear}.

For moderately small vibration amplitudes, this reduces to the well-known Duffing model~\cite{Duffing1918, Aldridge2005}, where the leading nonlinear contribution arises from the quartic term $\gamma_4$. At larger amplitudes, higher-order even nonlinearities become increasingly relevant~\cite{Miller2021, Huang2019, Kacem2009, polunin2016characterization, Kaisar2022, Samanta2018}. When spatial symmetry is broken, for instance, due to a static deflection of the structure, odd-order nonlinear coefficients $\gamma_{2i-1}$ also need to be considered. Their leading contribution is the cubic (Helmholtz) nonlinearity $\gamma_3$~\cite{Benedettini1987, eichler2013symmetry, Nabholz2020, ochs2021resonant, keskekler2022symmetry}.

One of the most common methods to extract nonlinear parameters is based on measuring the response of the system while a sinusoidal driving force of varying frequency is applied. For a simple Duffing resonator, this allows to extract the Duffing nonlinearity $\gamma_4$ quite reliably in a single-parameter fit. However, this requires precise knowledge of the eigenfrequency. The presence of temperature drifts and ubiquitous frequency fluctuations, a common scenario in the study of nanomechanical resonators, prevents the use of a pre-determined value for the eigenfrequency in the Duffing fit, as small shifts in eigenfrequency can significantly impact the resulting value of the Duffing nonlinearity. On the other hand, a simultaneous fit of the Duffing nonlinearity and the eigenfrequency can lead to ambiguity and yield inaccurate results. As soon as higher-order nonlinearities come into play, the situation becomes even more challenging, such that finite-element based frameworks have been used for their quantification~\cite{Keskekler2023}. 

Here, we demonstrate a method for precisely characterizing the nonlinear dynamics of a driven resonator. We focus on the regime in which the oscillation amplitude exhibits exponential decay, so that the action is simply related to the amplitude and nonlinear damping is negligible over the measured amplitude range. Although the conservative nonlinearities are small compared with the eigenfrequency, they have a pronounced effect on the driven response because of the low dissipation rate. By measuring the backbone curve, we systematically extract the dependence of the instantaneous oscillation frequency on the oscillation amplitude. This approach enables the characterization of resonators with arbitrary conservative nonlinearities, including the symmetry-breaking nonlinearities that have been disregarded in previous works ~\cite{polunin2016characterization,londono}. Further, it minimizes measurement uncertainties arising from the measurement electronics and thermal drifts.

The paper is structured as follows: In section~\ref{sec_setup}, a general description of the system is given along with its characterization in the linear and Duffing regime. The backbone method is introduced in section~\ref{sec_backbone} as an improved approach for the extraction of the nonlinear parameters. Lastly, section~\ref{sec_higher_order} presents an investigation of higher-order nonlinearities using the backbone along with some further observations.

\section{Setup and Duffing regime}
\label{sec_setup}
\begin{figure*}[hbtp!] 
\includegraphics[]{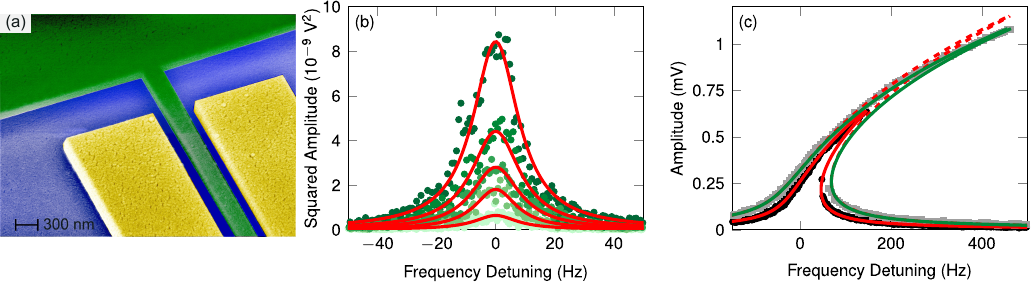}
\caption{(a)~Scanning electron micrograph of a doubly-clamped silicon nitride string resonator (green) and flanking electrodes (gold). ~(b)~Frequency response for different drive powers in the linear regime (\SI{-63}{\dBm}, \SI{-59}{\dBm}, \SI{-57}{\dBm}, \SI{-54.5}{\dBm}, \SI{-52}{\dBm} depicted as light to dark green dots). Red lines represent Lorentzian fits to the data. ~(c)~Frequency response for drive powers of \SI{-35}{\dBm} (black dots) and \SI{-30}{\dBm} (grey squares). The red solid line depicts a fit of the \SI{-35}{\dBm} curve with the Duffing model yielding $\alpha_{1}= \SI{4.170e25}{\per\kg\per\meter\squared}$, see Eq.~(\ref{eq_driven_response_duffing}). The red dashed line shows the same model at \SI{-30}{\dBm}. The green solid line shows a fit with the extended model with $\alpha_{2} = \SI{8.402e46}{\second\per\kg\squared\per\meter\tothe{4}}$.}
\label{fig:cal}
\end{figure*}

The system of interest consists of a nanomechanical doubly-clamped string resonator, similar to the one depicted in Fig.~\ref{fig:cal}(a). It is fabricated from strongly pre-stressed stoichiometric silicon nitride on a fused silica substrate, and features a width of \SI{270}{\nano\metre}, a thickness of \SI{100}{\nano\metre}, and a length of \SI{55}{\micro\metre}. Two adjacent electrodes enable dielectric drive and microwave cavity-assisted heterodyne detection~\cite{unterreithmeier2009universal,Faust2012,Rieger2012}. The microwave cavity is pumped on resonance at approximately \SI{3.6}{\giga\hertz} to enable precise displacement detection while avoiding any unwanted backaction effects~\cite{Faust2012}. An RF drive tone at $\omega_{\textsubscript{d}}$ is applied along with a DC voltage of \SI{5}{\volt}~\cite{unterreithmeier2009universal}. The DC voltage is selected to tune the fundamental out-of-plane mode of the nanostring to avoid hybridization with other modes, justifying the single-mode approximation applied in this work. Notice that the applied DC voltage also breaks the spatial symmetry of the electromechanical system~\cite{ochs2021resonant}. The experiment is performed at room temperature of \SI{293}{\kelvin}, and under high vacuum at a pressure around \SI{e-4}{\milli\bar}. 

The linear response of the system for a range of weak drive powers between \SI{-63}{\dBm} to \SI{-52}{\dBm} is depicted in Fig.~\ref{fig:cal}(b), along with the corresponding Lorentzian fits. We find a bare eigenfrequency $f_{0} = \omega_{0}/(2\pi)$ $\approx$ \SI{6.519}{\mega\hertz} and a full-width at half-maximum linewidth $2\Gamma/(2\pi) \approx$ \SI{21}{\hertz} with amplitude decay rate $\Gamma$, corresponding to a quality factor of approximately $300,000$. This data is used for calibration, as discussed in the App.~\ref{app:calibration} and described in previous work~\cite{huber2020spectral}. 

For stronger drive powers, the response becomes asymmetric as a consequence of the onset of nonlinearity, see Fig.~\ref{fig:cal}(c).
The asymmetric response curve indicates that the system enters the nonlinear regime where the vibration eigenfrequency depends on the amplitude of the system. 

Rather than parametrizing the conservative nonlinearity through the coefficients 
$\gamma_i$ of the potential expansion in Eq.~\eqref{eq:fullpot}, we work 
with the frequency--amplitude relationship
\begin{equation}\label{eq:omegaA}
    \omega(A) = \omega_0 + M\alpha_1\frac{A^2\omega_0}{2} 
              + M^2\alpha_2\frac{A^4\omega_0^2}{4} 
              + M^3\alpha_3\frac{A^6\omega_0^3}{8} + \ldots,
\end{equation}
which follows directly from expanding $\omega_\mathrm{I}(I) = \partial H/\partial I$ in powers 
of the action and substituting $I \propto A^2$. The quantity $\omega(A)$ denotes the nonlinear eigenfrequency of the freely vibrating resonator. It is equivalent to $\omega_\mathrm{I}(I)$ after expressing the action in terms of the
vibration amplitude.
Since the exact origin of the nonlinearities is not known, we can not map the coefficients $\alpha_i$ back to the $\gamma_i$. However, knowing the expansion coefficients of $\omega_\mathrm{I}(I)$ is sufficient to describe the driven response. This parametrization is both more 
natural and more practical: the instantaneous oscillation frequency 
$\omega(t)$  during the ringdown is related to $\omega(A)$, making the nonlinear frequency‑shift coefficients $\alpha_i$ ($i \geq 1$) directly and unambiguously accessible from experiment (see App.~\ref{app_theory} for more details). The $\gamma_i$, 
by contrast, enter $\omega(A)$ only indirectly --- through orbit-averaged integrals 
in which multiple coefficients contribute at each order in amplitude and cannot be extracted from the experiment without additional modeling assumptions. In this description, all other nonlinearities, including coupling to other modes, are also included.

In the following, we refer to $\omega(A)$ as the nonlinear eigenfrequency, and $\omega_0$ as the bare frequency of the (linear) system. At a moderately weak drive, the system is in the Duffing regime~\cite{nayfeh2024nonlinear}, where $\alpha_{2,3,...} = 0$.
For a spatially symmetric system with $\gamma_3=0$ one finds~\cite{kovacic}
\begin{equation}
    \alpha_{1} = \frac{3}{4M\omega^2_{0}}  \gamma_{{4}}.
\end{equation}
Breaking spatial symmetry leads to a modified  expression including the $\gamma_3$ parameter (Helmholtz-Duffing model) ~\cite{nayfeh2024nonlinear,ochs2022frequency,kovacic} %
\begin{equation}
    \alpha_{1} = \frac{3}{4M\omega^2_{0}}  \left(\gamma_{4} - \frac{10 \gamma^2_{{3}}}{9\omega^2_{0}} \right).
\end{equation}

To reliably characterize the nonlinear regime, we begin by probing the system at a drive power of \SI{-35}{\dBm}. The corresponding bidirectional frequency response is shown as black dots in Fig.~\ref{fig:cal}(c). At this drive power, a pronounced asymmetry emerges, enabling a meaningful extraction of the nonlinear response. A quantitative measure of the strength of the nonlinearity can be obtained from the saddle-node bifurcations. Specifically, we classify the response as sufficiently nonlinear when the upper bifurcation point occurs at a detuning $\delta\omega/(2\pi)$ of at least $10\,\Gamma/(2\pi)$.

The data is fit to the Duffing model (see Eq.~(\ref{eq_driven_response_duffing}) in App.~\ref{app_theory}) with $\alpha_{2,3,...} = 0$ and $\alpha_1$ as the only fitting parameter. The bare eigenfrequency is fixed at the value obtained from the linear characterization (see Fig.~\ref{fig:cal}(b)). The fit is included in Fig.~\ref{fig:cal}(c) as a red line. We find $\alpha_{1} = \SI{4.170e25}{\per\kg\per\meter\squared} $. 

While the fit of the response curve matches the data seemingly well, it does not yield an accurate value for the nonlinearity parameter. This is a consequence of a lack of precision in the input parameters. First and foremost, the bare eigenfrequency experiences drifts in the $\Gamma/(2\pi)$ range. These drifts occur in the course of a few response measurements, such that the eigenfrequency extracted from the linear characterization may differ from the one obtained in subsequent measurements in the nonlinear regime. 
Very small eigenfrequency drifts have a significant effect on the fit and thus the estimation of the nonlinearity as the frequency response scales inversely proportional to the difference between the nonlinear and bare eigenfrequency as seen in Eq.~(\ref{eq_driven_response_duffing}). This can be observed by repeating the fit after artificially shifting the resonance frequency, see App.~\ref{app:freqshift}. 
To mitigate the effect of eigenfrequency drifts, both the nonlinearity and the bare eigenfrequency can be used as fitting parameters. However, the two-parameter fit of the response yields equally large uncertainties due to the increased number of fitting parameters as quantified in App.~\ref{app:twofitsweep}.    

\section{Backbone method}
\label{sec_backbone}

\begin{figure*}
\includegraphics[]{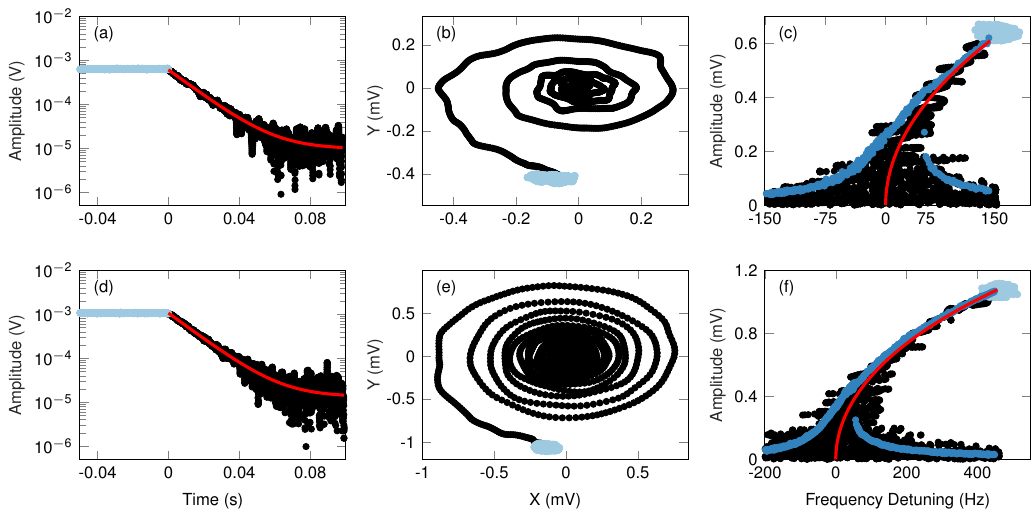}
\caption{Ringdown measurements in and beyond the Duffing regime. Turquoise and black dots show the response amplitude before and after switching off the drive. (a),(d)~Amplitude response at a drive power of \SI{-35}{\dBm} ( \SI{-30}{\dBm}) applied at the upper bifurcation point. The red line represents an exponential fit to the data. ~(b),(e)~Phase-space representation of the ringdown measurement, depicting the quadrature (Y) vs. the in-phase (X) component for the data shown in (a) and (d), respectively. The spiraling trajectory confirms the decay of the amplitude.~(c),(f)~Backbone relation for the ringdowns shown in (a) and (d), respectively. The backbone is extracted from the time derivative of the phase signal. It is shifted in frequency by the fitted detuning $\delta\omega/(2\pi)$ to be represented as a frequency difference from $\omega_0/(2\pi)$. The red lines show the fits of each backbone.
The frequency response measured at the respective drive power is superimposed on the backbone data as blue dots. It is plotted as a function of the frequency detuning $\delta\omega/(2\pi)$ from the bare eigenfrequency. 
}
\label{fig:-35_-30}
\end{figure*}

In this paper, we present an approach to precisely determine the nonlinear parameters of a nanomechanical system based on the ringdown method, thereby avoiding all the complications described previously. 
It relies on a single ringdown measurement at the upper bifurcation point of the response curve. This requires a careful initialization of the system. To this end, we start with a sweep at a low drive power to obtain the linear response of the system. 
Then, the drive power is increased to the desired value, where a bidirectional sweep is recorded to identify the bifurcation points. Lastly, the frequency is swept upwards from below the resonance to follow the upper amplitude branch until the selected frequency is reached. If not stated otherwise, we always consider the situation where this is the frequency of the upper bifurcation point where $A = A_{\textsubscript{max}}$, i.e.
where the drive frequency $\omega_{\textsubscript{d}}$ is set to $\omega_{\textsubscript{max}}=\omega(A_{\textsubscript{max}})$. 
The drive power remains on for at least $10/\Gamma$, which in our case corresponds to \SI{0.1}{\second} to make sure that the steady state is reached. Then the drive is turned off while the decaying amplitude signal is recorded over time for at least 10/$\Gamma$. Here we chose a recording time of 15/$\Gamma$ $\approx$ \SI{0.2}{\second}.

All measurements are performed on an HF2LI lock-in amplifier by Zurich Instruments. It is used to demodulate the signal at the drive frequency $\omega_{\textsubscript{d}}$, and thus provides the envelope of the oscillatory motion. We record both the in-phase and the quadrature component of the envelope signal. This is a key aspect of our measurement approach as it reveals both the instantaneous frequency and phase. Note that the demodulator’s bandwidth must be set large enough to capture the complete energy of the resonator as it rings down the backbone curve~\cite{Antoni2012,polunin2016characterization, catalini2020soft}.

As a first step, we compute the action directly from the measured amplitude and the inferred instantaneous frequency, and evaluate the ringdown of the action as a function of time. 
Even for the largest amplitude explored in this work, the action ringdown is exponential and can be fit with an energy decay time constant of \SI{8}{\milli\second}, in good agreement with $1/(2\Gamma)$ for $2\Gamma/(2\pi) \approx$~\SI{21}{\hertz} that was obtained by fitting the linear response to a Lorentzian lineshape (see App.~\ref{app:action} and Fig.~\ref{fig:actiondecay} for details). This experimentally justifies the use of the amplitude as the more intuitive, directly observable quantity for the following analysis.

Fig.~\ref{fig:-35_-30}(a)-(c) depicts the data obtained from the ringdown measurement at \SI{-35}{\dBm} whose frequency response is seen in Fig.~\ref{fig:cal}(c). In Fig.~\ref{fig:-35_-30}(a) the amplitude decay is plotted over time. The constant amplitude during driven vibration (turquoise dots) starts decaying after the drive is switched off at time $t_0 = 0$\,s (black dots). During the ringdown, the amplitude decays exponentially such that 
\begin{equation}
A(t) = A e^{-\frac{t}{ 2\tau}}
\label{eq:ring}
\end{equation}
where $\tau = 1/(2\Gamma)$ is the time constant of the energy decay. The exponential fit of Eq.~\ref{eq:ring} to the decaying amplitude (red line) yields $\tau =$~\SI{7.35}{\milli\second}. This is in good agreement with the $2\Gamma/(2\pi) \approx$~\SI{21}{\hertz} that was obtained by fitting the linear response to a Lorentzian lineshape. 

In Fig.~\ref{fig:-35_-30}(b) the same signal and color coding are displayed in a rotating phase space, namely, we show the quadrature signal against the in-phase signal. As long as the drive is on, the system performs forced vibrations with constant amplitude and phase, which in turn correspond to a stationary point in the rotating phase space. Due to fluctuations, this is seen as a cloud of points depicted in turquoise. Once the drive is turned off, the system performs a spiraling trajectory shown in black. Apart from the analysis of the backbone, which we explain in the following, the phase-space representation offers an additional advantage. It allows us to determine the exact start of the ringdown $t_{0}$ with higher accuracy than from the amplitude decay, see App.~\ref{app:ringstart} for details. 

Capturing both the in-phase and the quadrature component allows us to calculate the instantaneous phase $\phi$ as $\tan \phi = Y/X$~\cite{catalini2020soft}. Its derivative corresponds to the instantaneous frequency $\omega_{\textsubscript{inst}} = 2\pi d\phi/dt$, namely the demodulated frequency as the amplitude decays and the eigenfrequency moves along the backbone. The instantaneous frequency is described by
\begin{align}
    \omega_\mathrm{inst}(A) &= \omega(A) - \omega_{\textsubscript{d}} \,.
    \label{eq:ringfreq}
\end{align}
It is visualized in Fig.~\ref{fig:-35_-30}(c), where we plot the amplitude as a function of the instantaneous frequency. For the sake of clarity, the data is shifted by the detuning of the drive from the bare eigenfrequency, $\delta\omega = \omega_{\textsubscript{d}} -\omega_0$, such that the plot corresponds to $\omega(A)-\omega_0$. The turquoise dots show the amplitude in the presence of the drive. As soon as the drive is turned off, the system rings down, as shown by the black dots. As the detuning of the ringdown was set to the upper bifurcation point, the data captures the complete backbone curve for the chosen drive power. This is crucial for the correct determination of the nonlinearities. The backbone is fitted with Eq.~(\ref{eq:ringfreq}). In the Duffing regime discussed here, we assume that $\alpha_{2,3,...} = 0$, such that $\alpha_1$ and the detuning $\delta\omega = \omega_{\textsubscript{d}} -\omega_0$ are the only fit parameters. The value of $\omega_0$ is obtained from the characterization in the linear regime. The backbone fit is included in Fig.~\ref{fig:-35_-30}(c) as a red line, and yields $\alpha_{1} = \SI{4.097e25}{\per\kg\per\meter\squared}$
and an initial detuning $\delta\omega/(2\pi)$ of \SI{135}{\hertz}. 
This is in excellent agreement with the analysis of the Duffing frequency response, validating the quality of the backbone method for extracting nonlinear parameters. Notice, that the backbone decays to the bare eigenfrequency $\omega_0$, which appears shifted to \SI{0}{\hertz} in Fig.~\ref{fig:-35_-30}(c).  

The blue dots shown in Fig.~\ref{fig:-35_-30}(c) represent the previously recorded frequency response as a function of the detuning $\delta\omega$ at the same drive power. We observe that the phase-space cloud coincides with the upper bifurcation point of the response, thereby confirming that we indeed perform the ringdown at the upper bifurcation point as intended. 
 
Even though ringdown based approaches have been used before \cite{polunin2016characterization,catalini2020soft,londono}, our method can be applied without prior knowledge of the system's nonlinearities and requires only a record of the phase over time without the need for further processing apart from numerical derivation of the phase to obtain the backbone. It should be noted here that as the amplitude decays towards the noise floor, the (unwrapped) phase also becomes dominated by noise. As a result, the values of 
the instantaneous frequency given as the derivative of the phase also scatter across the frequency axis. Therefore, it is recommended to use a weighted fit with weights corresponding to the square of the amplitude to ensure unbiased parameter estimation by minimizing the effect of the noise on the fit.

\section{Higher order nonlinearities}
\label{sec_higher_order}

For stronger drives, the response of the system goes beyond the Duffing regime, and higher order nonlinearities must be included in the description~\cite{ochs2021resonant}. This is apparent in the frequency response shown in Fig.~\ref{fig:cal}(c) for a drive power of \SI{-30}{\dBm} (gray squares). The red dashed line depicting the expected behavior within the Duffing regime, i.e., the response calculated for the previously determined value of $\alpha_{1}$, clearly deviates from the data. To quantify the deviation, we use the normalized difference between $\omega_{\textsubscript{max}}$ calculated with the Duffing model and its measured value, $\lvert \omega_{\textsubscript{max}}^{\textsuperscript{th}}-\omega_{\textsubscript{max}}^{\textsuperscript{exp}}\rvert / (\omega_{\textsubscript{max}}^{\textsuperscript{th}} - \omega_0)$. If this ratio exceeds $5$\%, we include the next order of $\alpha$ to the nonlinear model.
This threshold is empirical, but applicable for our case, as it corresponds to a deviation on the order of the resonator's linewidth. 

The green solid line in Fig.~\ref{fig:cal}(c) shows a fit of the measured response with Eq.~(\ref{eq_driven_response}) including the $\alpha_{2}$-term. Using the previously determined value of $\alpha_{1}$ as a fixed input parameter, we obtain $\alpha_{2} = \SI{8.402e46}{\second\per\kg\squared\per\meter\tothe{4}}$.
The extended nonlinear model shows excellent agreement with the experimental data. For even stronger drives, similar metrics can be defined, allowing the determination of the onset of even higher-order nonlinear terms in the description of the response curves, facilitating the accurate description of the system.

For a more complete analysis, we perform a ringdown measurement at the same drive power of \SI{-30}{\dBm} and repeat the previously described procedure. Consistent with the analysis of the response curve, we include the $\alpha_{1}$ and $\alpha_{2}$ nonlinearities in the description of the backbone relation in Eq.~(\ref{eq:ringfreq}). Figure~\ref{fig:-35_-30}(d)-(f) presents the amplitude signal as well as the corresponding phase space representation and backbone during the ringdown. As in (a)-(c), the data taken before and after switching off the drive are shown in turquoise and black, respectively. Fitting the backbone equation with the previously determined value of $\alpha_{1}$ as a fixed input parameter, we obtain a detuning of \SI{459}{\hertz} and $\alpha_{2} = \SI{8.4162e46}{\second\per\kg\squared\per\meter\tothe{4}}$.

\begin{figure}[htbp!]

\includegraphics[width=\columnwidth]{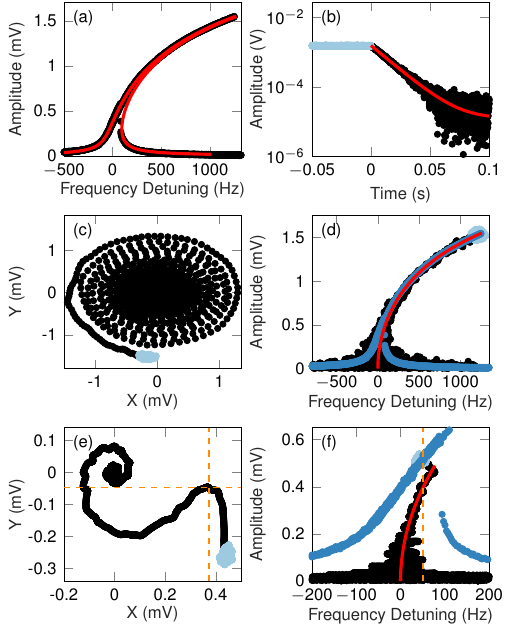}

\caption{(a) Frequency response at a drive power of \SI{-26}{\dBm}. The red solid line shows a fit with the extended nonlinear model that includes $\alpha_2$ and $\alpha_3$. 
~(b)~Ringdown measurement for the same drive power applied at the upper bifurcation point. Turquoise (black) points show the amplitude signal before (after) switching off the drive. The red line represents the exponential fit to the data. ~(c)~Phase space representation of the ringdown measurement, showing the quadrature (Y) vs. the in-phase (X) component of the data using the same color code as in~(b).~(d)~Backbone relation for the ringdown. The backbone is extracted from the time derivative of the phase signal. 
It is shifted in frequency by the fitted detuning $\delta\omega/(2\pi)$ to be represented as a frequency difference from $\omega_0/(2\pi)$. 
The frequency response measured at the same drive power is superimposed
to the backbone data as blue dots.
It is plotted as a function of the frequency detuning $\delta\omega/(2\pi)$ from the bare eigenfrequency. This allows to overlay the backbone and the swept response data on the same frequency axis.
~(e)~Phase space trajectory of a ringdown measurement at a drive power of \SI{-26}{\dBm} and a detuning of \SI{50}{\hertz}. Turquoise (black) points show the amplitude signal before (after) switching off the drive. The orange-dashed cross indicates the point where 
the trajectory reverses its rotational direction as the instantaneous frequency crosses the reference frequency.~(f)~Backbone relation for the ringdown in~(e). The backbone is extracted from the time derivative of the phase signal, and plotted shifted by $\delta\omega/(2\pi)$. The orange-dashed line illustrates the reference frequency. The red line shows a fit of the backbone.  
The frequency response as a function of $\delta\omega/(2\pi)$ measured at the same drive power is superimposed on the backbone data as blue dots.}
\label{fig:-26ring}
\end{figure}

To prove the generality of the backbone method as a means of determining nonlinearities accurately, we perform a frequency response and ringdown measurement at an even higher drive power of \SI{-26}{\dBm} shown in Fig.~\ref{fig:-26ring}(a)-(d). Using the same criterion for determining when to include higher order nonlinearities (in this case using $\omega_{\textsubscript{max}}^{\textsuperscript{th}}$ computed including $\alpha_1$ and $\alpha_2$), we observe that the contribution from $\alpha_3$ needs to be taken into account. We perform the response fit keeping the existing coefficients constant, but with an additional coefficient obtaining 
$\alpha_{3} = \SI{6.176e68}{\second\squared\per\kg\tothe{3}\per\meter\tothe{6}}$. 

Again, we also use the backbone technique, which yields
$\alpha_{3} = \SI{5.783e68}{\second\squared\per\kg\tothe{3}\per\meter\tothe{6}}$
and a detuning of \SI{1208}{\hertz}.
We wish to note that despite the strong impact of higher-order nonlinearities, the overall effect of these nonlinearities is still weak, i.e. $\omega_0 \gg M^{i} \alpha_i A^{2i} \omega_0^i/2^i$ for all observed $i$.

As previously discussed (see also App.~\ref{app:freqshift} and Fig.~\ref{fig:freqshift} therein), the nonlinearity parameters extracted from the response curve critically depend on the exact value of the bare eigenfrequency. In the driven response equation (Eq.(~\ref{eq_driven_response_duffing})), $\omega_0/(2\pi)$ appears in the resonance term additively and multiplicatively in the amplitude scale factor $\frac{F_\mathrm{d}^2}{4\omega_0^2M^2}$. Consequently, an error in $\omega_0/(2\pi)$ forces the fit to artificially rescale the nonlinear coefficients to compensate for the frequency shift. By contrast, the instantaneous frequency from which we fit the backbone is proportional to the bare resonance frequency $\omega_0/(2\pi)$. Even though we use the value from the linear response for $\omega_0/(2\pi)$, we account for static drift-induced shifts with an additional fitting parameter $\delta\omega/(2\pi)$ that enters the relation only additively. By fitting this detuning, we essentially correct and absorb any possible frequency drifts that could have occurred between measurements, and we can successfully mitigate their effect. 

Another frequently overlooked aspect is the difference in measurement timescales. Data acquisition for a frequency response needs to include sufficient settling times between discrete points to ensure a steady-state response, resulting in extended measurement durations for high-Q resonators. Over this long acquisition window, $\omega_0$ experiences dynamic drift. On the other hand, a single ringdown trace can be acquired much faster and is thereby shielded against drifts that occur during the response measurement.

To evaluate the sensitivity of frequency-response analysis to frequency drifts accumulated during the measurement, we also performed fits in which the detuning was treated as a second free parameter. Therefore, to further validate our method and to quantify the robustness of the backbone method, we also compared frequency response fits of the $\alpha$ coefficient with and without this additional detuning. The inclusion of a detuning parameter in the frequency response fit increases the uncertainty of the extracted nonlinear coefficients and can significantly alter their values, as shown in App.~\ref{app:twofitsweep}. Even though in the backbone method, a detuning is fitted simultaneously, it yields for our measurements, stable nonlinear coefficients with considerably smaller error (below $10\%$). This is a very important outcome, as the determination of the nonlinear coefficients remains largely insensitive to resonance-frequency drifts. Although the detuning is fitted simultaneously, the extracted nonlinear parameters remain robust as the ringdown captures in a single fast (relative to the frequency response) trace the entire amplitude range down to the noise floor. Our method can therefore be used to accurately determine nonlinearities while strongly suppressing the effect of static and dynamic frequency shifts.

The use of the ringdown method to investigate the dynamics of the system led to another interesting observation. For the ringdown, the system is driven at a specific frequency, which also serves as the reference frequency based on which the lock-in amplifier demodulates the signal. For the data discussed so far, the reference frequency was chosen as the upper bifurcation point. Then, the instantaneous frequency is continuously reduced as the amplitude decreases during the ringdown. This is seen as a decay down the backbone curve until it reaches the bare eigenfrequency, c.f. Figs.~\ref{fig:-35_-30}(c),(f) and Fig.~\ref{fig:-26ring}(d). 
If, however, the reference frequency is chosen below the upper bifurcation point, the instantaneous frequency will jump to the same amplitude on the backbone curve once the drive is off. Only then will it decay down the backbone curve as in the previous situation~\cite{Antoni2012, polunin2016characterization, Guettinger2017}. 
For initial driving frequencies between $\omega_0$ and $\omega_{\textsubscript{max}}$, the decaying instantaneous frequency crosses the reference frequency. At this point, the trajectory in phase space changes its sense of rotation. 
This is observable for small detunings and moderate drive powers, measured at a high enough sampling rate.

Figure~\ref{fig:-26ring}(e) and (f) show an example for a ringdown performed at a drive of \SI{-26}{\dBm} with a detuning of \SI{50}{\hertz} from the bare resonance frequency. The phase-space representation of the ringdown in Fig.~\ref{fig:-26ring}(e), plotted with the same color code as before, clearly reveals the reversal in the trajectory's direction of rotation from counterclockwise to clockwise, indicated by an orange-dashed cross.
The crossing point ($0.37$\,mV/$-0.05$\,mV) corresponds to an amplitude of \SI{0.373}{\milli\volt}.

The corresponding backbone curve in Fig.~\ref{fig:-26ring}(f) corroborates this finding. There is a distance of a few \SI{10}{\hertz} between the phase-space cloud of the driven response (turquoise) and the parabolic backbone curve, heralding the jump of the instantaneous frequency to the backbone thanks to the large sampling rate employed in the measurement. 
As also seen in Fig.~\ref{fig:-26ring}(f), the amplitude decaying down the backbone crosses the reference frequency marked by an orange-dashed line at an amplitude of approximately \SI{0.4}{\milli\volt}. This coincides with the amplitude of the phase space trajectory at the point marked by the orange-dashed cross in Fig.~\ref{fig:-26ring}(e). 

\section{Conclusion}
\label{sec_conclusion}
We present a framework for precisely and robustly extracting the conservative nonlinearities of a nonlinear nanomechanical resonator. The method relies on the backbone curve obtained from a ringdown measurement. We study the response of a high-Q nanomechanical string resonator to demonstrate the performance of the backbone method. By comparing its results with the established nonlinear characterization via the frequency response curve, we demonstrate both the accuracy and the superiority of the backbone method over existing schemes. 

Conventionally, conservative nonlinearities are determined by fitting the frequency response curve obtained at a large drive power. Even though this method is widespread, it suffers from inaccuracies due to drifts of the bare eigenfrequency over the measurement period. 

The backbone method relies on a ringdown measurement that simultaneously records the in-phase and quadrature signal. This allows us to access the instantaneous frequency of the system~\cite{catalini2020soft}. This frequency is described by the backbone relation and can be used to fit nonlinearities without relying on an accurate determination of the bare eigenfrequency. We show excellent agreement between nonlinear coefficients determined using both methods, but also clearly reveal how small frequency drifts limit the accuracy of the frequency response analysis, whereas the ringdown remains unaffected by changing measurement conditions as the system freely decays. This is also a consequence of the difference in measurement timescales, as a single ringdown trace is acquired much faster than a frequency response. This identifies the backbone technique as the method of choice for an accurate determination of nonlinearities.

We emphasize that the backbone method is not limited to non-symmetry-breaking nonlinearities, as literature on the direct analysis of the ringdown curve~\cite{polunin2016characterization,londono}, but explicitly includes symmetry-breaking terms. Moreover, the backbone technique can be extended to include dissipative nonlinearities as they manifest as a non-exponential amplitude decay~\cite{polunin2016characterization}. As dissipative nonlinearities were negligible for the drive power range studied in this work, this remains a topic for future research.
The same applies to even more nonlinear regimes of nonsinusoidal oscillations.

\begin{acknowledgments}
We gratefully acknowledge financial support from the Deutsche Forschungsgemeinschaft (DFG, German Research Foundation) through Project-ID No.425217212-SFB 1432 and under Germany’s Excellence Strategy—EXC-2111—390814868. The research is further supported by the Bavarian state government with funds from the Hightech Agenda Bavaria. We would also like to thank Mark Dykman for the insightful discussions and his useful comments.

\end{acknowledgments}

\appendix
\section*{APPENDIX}

\section{Theory}
\label{app_theory}
In this Appendix, we provide a brief overview of the theoretical description of the mode dynamics. As explained in the main text, generally speaking, many modes of a nonlinear system may couple to each other. However, if one specific mode is driven and the others are off resonant, only one mode is excited to large amplitudes. We approximate the dynamics of this mode by a resonantly driven oscillator with position $q$ and momentum $p$ that moves in the potential $U(q)$ from Eq.~(\ref{eq:fullpot}). The dynamics are then described by
\begin{align}
    \dot{q} &= \frac{\partial H(q,p,t)}{p}, & \dot{p} &= - \frac{\partial H(q,p,t)}{q} - 2 \Gamma p. \label{eq_motion_q_p}
\end{align}
Here, $2 \Gamma$ is the energy decay rate of the mode, and the Hamiltonian is given by
\begin{align}
    H(q,p) &= \frac{p^2}{2M} + U(q) - F_\mathrm{d} \cos(\omega_\mathrm{d} t),\label{ham}
\end{align}
where $M$ the effective mass and $F_\mathrm{d}$ and $\omega_\mathrm{d}$ are the amplitude and the frequency of the drive, respectively. In the following, we are interested in the regime where the drive amplitude $F_\mathrm{d}$ is small, such that $U(q)$ is dominated by the quadratic contribution proportional to the bare frequency $\omega_0$. This is the case for weak nonlinearities and therefore also small frequency shifts, i.e. $|\omega(A) - \omega_0| << \omega_0$. Further, we assume that $\Gamma/\omega_0 \ll 1$ such that the nonlinearities still affect the response as soon as  $|\omega(A) - \omega_0| > \Gamma $. As a first step, we look at the isolated dynamics of the mode.

\subsection{Isolated Mode}
If the mode is isolated, i.e., $F_\mathrm{d} = \Gamma = 0$, the energy is conserved. The system is best described using so-called action-angle variables defined by~\cite{Landau1976} 
\begin{equation}
I = (2\pi)^{-1} \oint p \ dq  \quad \text{and} \quad  \psi = \partial_{I} \int p \ dq \label{eq:action_angle}.
\end{equation}
An isolated mode performs periodic vibrations with $I = \text{const.}$. The phase linearly accumulates in time, $\psi = \omega_\mathrm{I}(I) t$. The vibration frequency is $\omega_\mathrm{I}(I) = \mathrm{d} E/\mathrm{d} I$, where $E$ is the mode energy. 

For a weakly nonlinear mode, where $U(q)$ is dominated by the term quadratic in $q$, we can expand the vibration frequency in powers of the action variable:
\begin{align}
    \omega_\mathrm{I}(I) = \omega_0 + \alpha_1 I + \alpha_2 I^2 + \alpha_3 I^3 + \ldots \,.
\end{align}
The relation between energy and action variable can be made explicit by noting that
\begin{align}
    E = \int_0^I \omega_\mathrm{I}(I') \mathrm{d} I' = \omega_0 I + \frac{\alpha_1}{2} I^2 + \frac{\alpha_2}{3} I^3 + \ldots .
\end{align}
In order to relate the coefficients $\alpha_i$ with the potential $U(q)$, see for example \cite{ochs2022frequency}, where it was found that
\begin{align}
&\alpha_{1} = \frac{3}{4M\omega^2_{0}}  \left(\gamma_{4} - \frac{10 \gamma^2_{3}}{9\omega^2_{0}} \right),\label{eq:alpha_coefficients}\\
&\alpha_{2} = \frac{5\gamma_{6}}{4M^2\omega^3_{0}}-\frac{7\gamma_{3}\gamma_{5}}{2M^2\omega^5_{0}}-\frac{51\gamma^2_{4}}{64M^2\omega^5_{0}} + \frac{75\gamma^2_{3}\gamma_{4}}{16M^2\omega^7_{0}}-\frac{235\gamma^4_{3}}{144M^2\omega^9_{0}}. \nonumber
\end{align}
Note that generally speaking, the functional dependence of the vibration frequency on the action may also depend on coupling to other modes. Here, we only discuss this simple model of a nonlinear potential as one possible source of nonlinearities. The expansion of the vibration frequency in terms of the parameters $\alpha_i$ also holds in more general cases. It is only limited by the constraint that the action remains small.

\subsection{Damped Mode}
We now study the dynamics of the mode under the influence of dissipation. Then the energy is no longer conserved, and similarly the action becomes time dependent. Importantly, in the regime where $\Gamma /\omega_0 \ll 1$, the action $I$ only decays slowly, while the phase $\psi$ performs fast oscillations with the frequency $\omega_\mathrm{I}(I)$ being approximately constant over one period. Averaging over these fast oscillations results in \cite{polunin2016characterization} 
\begin{align}
    \dot{I}(t) &= - 2\Gamma I(t), & \dot{\psi}(t) &= \omega_\mathrm{I}(I(t)),
\end{align}
i.e., the action variable decays exponentially. 

In the main text, the instantaneous frequency $\omega_\mathrm{inst}$ was observed as a function of the amplitude of vibrations at the main tone. During the ringdown, the mode vibrates at different frequencies, and therefore, the definition of the amplitude is also ambiguous. In the experiment, the demodulator bandwidth is chosen sufficiently large such that the energy of the resonator during the ringdown around the main tone and the entirety of the backbone is captured. As also explained in the main text, in the regime of interest where the effect of nonlinearities is weak, the vibrations can be approximated as sinusoidal. Therefore, higher overtones can be neglected. To the lowest order in the amplitude, the relation to the action is given by 
\begin{equation}
    A \approx \sqrt{2I/M \omega_0}.
    \label{action_eq}
\end{equation}

Therefore, the amplitude decays exponentially with a time constant given by $\Gamma$, and the frequency of vibrations with a certain amplitude is given by
\begin{align}
    \omega (A) &= \omega_0 + M\alpha_{1}\frac{A^2\omega_{0}}{2} + M^2\alpha_{2}\frac{A^4\omega^2_{0}}{4}+ M^3\alpha_{3}\frac{A^6\omega^3_{0}}{8} + \ldots
\end{align}
which coincides with Eq.~(\ref{eq:omegaA}).

\subsection{Driven Mode}
We now consider the case where the oscillator is driven at frequency $\omega_\mathrm{d}$ with an amplitude $F_\mathrm{d}$. In action-angle variables Eq.~(\ref{eq_motion_q_p}) takes the form 
\begin{align}
    \dot{\psi} &= \omega_\mathrm{I}(I) - \frac{\partial q}{\partial I} (F_\mathrm{d} \cos(\omega_\mathrm{d} t) - 2 \Gamma p),\\
    \dot{I} &= \frac{\partial q}{\partial \psi} (F_\mathrm{d} \cos(\omega_\mathrm{d} t) - 2 \Gamma p).
\end{align}
The driven mode is expected to perform vibrations at the drive frequency with a constant amplitude $A_0$. We make the Ansatz
\begin{align}
    q = A_0 \cos (\psi(t)),
\end{align}
with $\psi(t) = \omega_\mathrm{d} t + \phi$ and seek the stationary solutions for the phase $\phi$ and amplitude $A_0$. With that, we find the previously derived result for the amplitude of forced vibrations \cite{bogoliubov,ochs2022frequency, ochs2021resonant}
\begin{align}
\label{eq_driven_response}
    \frac{F_\mathrm{d}^2}{4 \omega_0^2 M^2} &= A_0^2 \left[ (\omega(A_0) - \omega_\mathrm{d})^2 + \Gamma^2 \right].
\end{align}
For $\alpha_i = 0$ for $i>1$ this yields the standard result of a driven Duffing resonator~\cite{nayfeh2024nonlinear} 
\begin{align}
\label{eq_driven_response_duffing}
    \frac{F_\mathrm{d}^2}{4 \omega_0^2 M^2} &= A_0^2 \left[ (\omega_0 - \omega_\mathrm{d} + M \alpha_1 A_0^2 \omega_0 /2)^2 + \Gamma^2 \right].
\end{align}
\section{Action Ringdown}
\label{app:action}
As mentioned before, the assumption that the amplitude decays exponentially holds only in the case where the coupling to other modes is negligible, and the amplitude can then be calculated from the action via the simple relation in Eq.~(\ref{action_eq}). 

To verify that this assumption applies to our data, we computed the action directly from the action integral by calculating the area enclosed by each turn of the ringdown trajectory in phase space and relating it to the action.

\begin{figure}[hbt]
    \centering
    \includegraphics[]{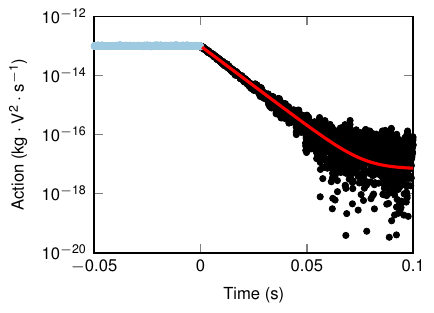}
    \caption{Action decay measured at the upper bifurcation point for a drive power of \SI{-26}{\dBm}, corresponding to the largest amplitude explored in this work. The ringdown follows an exponential decay, confirming the linear damping hypothesis.}
    \label{fig:actiondecay}
\end{figure}

The action ringdown from the upper bifurcation point at a drive power of \SI{-26}{\dBm} is shown in Fig.~\ref{fig:actiondecay}. This corresponds to the largest amplitude explored in this work.
The decaying action can be fitted with an exponential decay with an energy decay time constant of \SI{8}{\milli\second}, in excellent agreement with the energy decay constant of $1/(2\Gamma)$ extracted from $2\Gamma/(2\pi) \approx$~\SI{21}{\hertz} that was obtained by fitting the linear response to a Lorentzian lineshape.  
The action for lower drive powers decays exponentially with the same decay time as well. So, both action and amplitude decays are exponential, then the action-amplitude mapping is adequately described by Eq.~(\ref{action_eq}) without containing higher order terms. This convincingly demonstrates the sinusoidal character of the oscillations as well as the absence of nonlinear damping, thereby justifying the use of the more intuitive, experimentally observable amplitude $A$ rather than the action $I$. 
In the main text, we therefore only depict the ringdown of the amplitude.

\section{Calibration}
\label{app:calibration}

The data is calibrated in measured units of volts, employing the calibration method described in Ref.~\cite{huber2020spectral}. The sample is first actuated at low drive power to obtain linear frequency responses. At these low drive powers, the response amplitude is small, and therefore, the contribution of conservative and dissipative nonlinearities can be neglected. By varying the frequency of the RF driving tone, we can trace the response amplitude of the resonator. In the limit of weak damping (2$\Gamma \ll \omega_{0}$) and near resonance, the frequency response of the intensity is described by a Lorentzian lineshape
\begin{equation}
A^2 = \frac{ F_{\textsubscript{d}}^2}{4M^2\omega_{0}^2} \cdot \frac{1}{\delta\omega^2 + \Gamma^2}\label{eq:lor}
\end{equation}
with $\delta\omega = \omega_{\textsubscript{d}} - \omega_{0}$ the frequency detuning from resonance. 

In the linear regime, we can assume that not only the drive but also the response signal (both measured in \unit{\volt}) are linearly proportional to the driving force (in \unit{\newton}) and the deflection amplitude (in \unit{\metre}), respectively. The relations $A = a \cdot V_{\textsubscript{out}}$ and $F_{\textsubscript{d}}/M = b \cdot V_{\textsubscript{in}}$ express mathematically our assumption with calibration factors $a$, and $b$, respectively, where $M$ is the mode's effective mass. 

\begin{figure}[hbtp!] 
\includegraphics[]{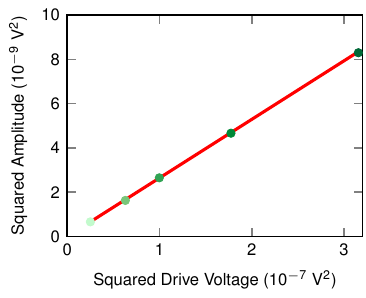}
\caption{ 
Extracted maximum of frequency response curves for drive powers ranging from \SI{-63}{\dBm} to \SI{-52}{\dBm} (light to dark green, as in Fig.~1(a) of the main text) plotted over the square of the driving voltage. From the linear fit, depicted in red, we can obtain the calibration parameter c that relates the drive to the response amplitude. In our case, we obtain $c = 0.2645$.}
\label{fig:cal_lin}
\end{figure}

We rewrite Eq.~(\ref{eq:lor}) in terms of frequency rather than angular frequency and express the amplitude and drive in units of \unit{\volt} as  
\begin{equation}
V_{\textsubscript{out}}^2 = \frac{1}{16\pi^4} \left(\frac{b}{a}\right)^2 \frac{1}{4f^2_{0}   {(\Gamma/(2\pi))}^2 } \frac{{(\Gamma/(2\pi))}^2}{\delta f^2 + {(\Gamma/(2\pi))}^2 } \cdot V_{\textsubscript{in}}^2.\label{eq:lorcal}
\end{equation}
We can then define a new dimensionless parameter $c = (b/a)^2 / (64\pi^4 f^2_{0} (\Gamma/(2\pi)^2)$, which we call the calibration factor. To obtain the calibration factor $c$, we record multiple responses at different drive powers in the linear regime by varying the frequency of the RF tone. The $c$ factor can be extracted either from each curve individually or by extrapolating the maximum of the response for each drive power and fitting a polynomial of degree one through the data as seen by Eq.~(\ref{eq:lorcal}). Figure~\ref{fig:cal}(b) depicts a set of Lorentzian frequency responses measured at different drive powers from \SI{-63}{\dBm} to \SI{-52}{\dBm} from which we obtain a resonance frequency around \SI{6.519}{\mega\hertz} and a linewidth 2$\Gamma/(2\pi)$ of \SI{21}{\hertz}. Figure~\ref{fig:cal_lin} shows the linear fit between the squared response amplitude (intensity) and squared drive voltage, yielding $c = 0.2645$. 

Another important parameter is the proportionality constant $a$, which can be directly determined from the thermomechanical motion of the resonator~\cite{hauer}. Since thermal noise is white, when it acts on the resonator, it results in a Lorentzian spectrum, which is described by 
\begin{equation}
    S_{xx}(\omega) = \frac{8\Gamma k_\mathrm{B} T}{M} \cdot \frac{1}{(\omega_0^2 - \omega^2)^2 + 4\Gamma^2\omega^2},
    \label{eq:thermomech}
\end{equation}
where $T$ is the temperature, $M$ the effective mass, $k_\mathrm{B}$ the Boltzmann constant.

For $|\omega-\omega_0|\ll \omega$, we can use the approximation $(\omega_0^2 - \omega^2) \approx 2\omega_0(\omega_0 - \omega)$ and Eq.~(\ref{eq:thermomech}) becomes 
\begin{equation}
    S_{xx}(\omega) \approx \frac{2\Gamma k_B T}{M\omega_0^2} \cdot \frac{1}{(\omega_0 - \omega)^2 + \Gamma^2}.
    \label{eq:thermomechapprox} 
\end{equation}

The last step is to convert $S_{xx}(\omega)$ given in \SI{}{\meter\squared\per\hertz} to our measured spectrum in units of \SI{}{\volt\squared\per\hertz}. For that, we use the factor $a$ so that $S_{VV}(\omega) =  \frac{1}{a^2} S_{xx}(\omega)$. 

\begin{figure}
    \centering
    \includegraphics[]{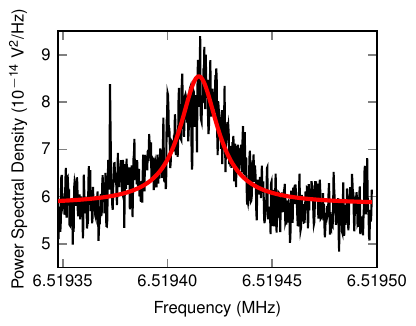}
    \caption{Measured thermomechanical spectrum of the fundamental mode. The data is depicted in black and the corresponding Lorentzian fit in red. From the fit, we can obtain the proportionality factor $a$.}
    \label{fig:thermomotion}
\end{figure}

The measured thermomechanical spectrum is shown in Fig.~\ref{fig:thermomotion}. It was obtained by averaging over $1,000$ traces since the signal is weak and buried in noise. A Lorentzian fit with only the calibration parameter $a$ as fit parameter in Eq.~(\ref{eq:thermomechapprox}) and assuming $T = \SI{293}{K}$, $M = \SI{2.11e-15}{\kilo\gram}$, $\omega_0 / (2\pi) = \SI{6.519}{\mega\hertz}$ and $2\Gamma/(2\pi) = \SI{21}{\hertz}$ is shown in Fig.~\ref{fig:thermomotion} as a red line. In our case, we find that $a = \SI{3.5181e-5}{\meter\per\volt}$. This allows us to quantify the remaining second calibration parameter $b$ as $ b = 2 \cdot a \cdot \sqrt{c} \cdot \omega_0 \cdot \Gamma$. We find $b \approx \SI{1.0245e5}{\milli\per\volt\per\second\squared}$.

Note that the data are recorded in terms of voltage and frequency, therefore all the fits are also performed in those units. With the help of the calibration factor $a$, however, we can also express the nonlinear coefficients $\alpha$ in SI units using the following conversion : $\alpha_i = p_i / (M^i a^{2i})$ where $p_i$ is the nonlinear coefficient in units of ${ ({s}^{i-1} {V}^{-2i})}$.

Notice that the determination of the $a$ factor by means of thermomechanical calibration represents a more direct approach than the one employed in Ref.~\cite{huber2020spectral}, where $a$ was obtained from the geometric nonlinearity of the sample. This was facilitated by improving the displacement sensitivity by means of an additional amplification stage, which now enables the detection of the thermomechanical motion of the resonator. Both calibration methods coincide.

\section{Effect of Small Resonance Frequency Shifts}
\label{app:freqshift}
\begin{figure}[htb!]
\includegraphics[width=\columnwidth]{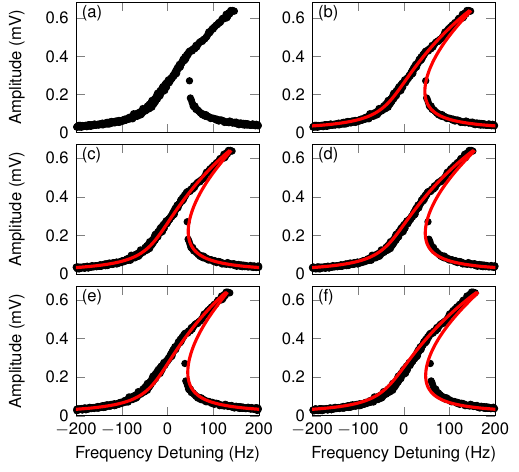}
\caption{Effect of frequency drifts on the response fit (a)~Bidirectional frequency response (black dots) at a drive of \SI{-35}{\dBm}.~(b)~Response fit of the data shown in (a) with unshifted resonance frequency. ~(c)-(d)~Response fit of the same data manually shifted by $\pm 1/2 \cdot \Gamma / (2\pi)$, namely \SI{5}{\hertz} and -\SI{5}{\hertz}, respectively.
~(e)-(f)~Response fits of the same data shifted by $ \Gamma / (2\pi) $, namely $\pm$\SI{10}{\hertz}. Responses shifted by $ \Gamma / (2\pi) $, namely $\pm$\SI{10}{\hertz} with $\alpha_{1}$ at \SI{3.676e25}{\per\kg\per\meter\squared} and \SI{4.595e25}{\per\kg\per\meter\squared}. 
The results of all fits are summarized in Tab.~\ref{tab:freqshifttab}
For small detunings, the fits show good agreement with the data. As the shift increases, the fits do not describe the data as accurately, and the obtained fitting parameter differs by $10\%$.}
\label{fig:freqshift}
\end{figure}

\begin{table}[b!]
\begin{ruledtabular}
\setlength{\tabcolsep}{8pt} 
\begin{tabular}{ccc}
Freq. Shift (\SI{}{\hertz}) & $\alpha_{1} (\SI{e25}{\per\kg\per\meter\squared}) $ & Coeff. Error (\%) \\
\hline
-10 & \num{4.595} & 10.7 \\
-5 &  \num{4.404} & 11\\
0 &  \num{4.170}  & 10.4\\
5 &  \num{3.944} & 11.4\\
10 & \num{3.676} & 10.16\\
\end{tabular}
\end{ruledtabular}
\caption{\label{tab:freqshifttab}
Values of the nonlinearity $\alpha_{1}$ obtained from fits to the frequency response discussed in Fig.~\ref{fig:freqshift}, along with the standard error. An artificial shift was added to the resonance frequency to simulate drifts that may occur between consecutive measurements. As the resonance decreases from its nominal value, $\alpha_{1}$ increases, whereas the opposite trend is observed with a positive frequency shift.}
\end{table}

A major advantage of using the backbone method to determine the conservative nonlinearities of the system is the mitigation of the effect of frequency drifts. To show how detrimental even small drifts (below the resonator's linewidth) can be on the nonlinearity parameter values, we artificially shifted the resonance frequency of the response curve measured at \SI{-35}{\dBm} by $\pm$\SI{5}{\hertz} and $\pm$\SI{10}{\hertz}. Figure~\ref{fig:freqshift}(a)-(b) depicts the original data as well as the correct Duffing fit. 
Figure~\ref{fig:freqshift}(c)-(d) as well as (e)-(f) show the response curve shifted by $\pm$\SI{5}{\hertz} as well as $\pm$\SI{10}{\hertz}, respectively, along with the corresponding fits. By visual inspection, it is hard to observe differences in the quality of the fits, especially for panels (c)-(d) featuring the smaller artificial shift. This indicates that small eigenfrequency drifts are not easily discerned in the data. Table~\ref{tab:freqshifttab} summarizes the obtained fitting values. It becomes clear that small drifts change the nonlinearity values by up to $10\%$ even though the quality of the fit remains comparable. The fitting parameter compensates for the drift by either increasing or decreasing its value from the true parameter.

This result indicates that even slight frequency drifts (below the linewidth) can change the fitting results. In standard measurement routines, the resonance frequency is first determined by a linear response sweep, and the drive is subsequently increased to obtain a nonlinear response. During the time between the two measurements, there can be slight drifts in the resonance frequency if no frequency stabilization mechanism is enabled, which, as shown here, can skew the fitting. Additionally, a wrongful assumption about the $\alpha_{1}$ parameter can also lead to an inaccurate determination of further nonlinearities.

\section{Parameter Fit and Error}
\label{app:twofitsweep}

\begin{table*}[t!]
\begin{ruledtabular}
\setlength{\tabcolsep}{8pt}
\begin{tabular}{ccccccc}
Parameter &
Method &
Fit. Par. &
Coefficient &
Detuning (Hz) &
Coeff. Error (\%) &
Detuning Error (\%) \\
\hline

$\alpha_{1}$ [\SI{}{\per\kg\per\meter\squared}]
& Response
& 1
& \num{4.170e25}
& Fixed
& \num{10.7} 
& --- \\

$\alpha_{1}$ [\SI{}{\per\kg\per\meter\squared}]
& Response + Drift
& 2
& \num{2.096e25}
& \num{45.39}
& \num{20.7}
& \num{9.65} \\

$\alpha_{1}$ [\SI{}{\per\kg\per\meter\squared}]
& Backbone 
& 2
& \num{4.097e25}
& \num{135.03}
& \num{1.7}
& \num{0.59} \\

\hline

$\alpha_{2}$ [\SI{}{\second\per\kg\squared\per\meter\tothe{4}}]
& Response
& 1
& \num{8.406e46}
& Fixed
& \num{12.6}
& --- \\

$\alpha_{2}$ [\SI{}{\second\per\kg\squared\per\meter\tothe{4}}]
& Response + Drift
& 2
& \num{9.230e45}
& \num{28.44}
& \num{118.1}
& \num{7.88} \\

$\alpha_{2}$ [\SI{}{\second\per\kg\squared\per\meter\tothe{4}}]
& Backbone 
& 2
& \num{8.419e46}
& \num{458.81}
& \num{8.3}
& \num{0.35} \\

\hline

$\alpha_{3}$ [\SI{}{\second\squared\per\kg\tothe{3}\per\meter\tothe{6}}]
& Response
& 1
& \num{6.277e68}
& Fixed
& \num{3.47}
& --- \\

$\alpha_{3}$ [\SI{}{\second\squared\per\kg\tothe{3}\per\meter\tothe{6}}]
& Response + Drift
& 2
& \num{8.983e68}
& \num{-53.32}
& \num{2.35}
& \num{4.22} \\

$\alpha_{3}$ [\SI{}{\second\squared\per\kg\tothe{3}\per\meter\tothe{6}}]
& Backbone 
& 2
& \num{5.783e68}
& \num{1208.05}
& \num{4.26}
& \num{2.2} \\

\end{tabular}
\end{ruledtabular}
\caption{\label{tab:fitcomparison}
Coefficient estimation extracted from the frequency response with and without an additional detuning parameter to account for frequency drifts, and from the two-parameter backbone fit. Compared to the single parameter fit from the response curve, the backbone method results in substantially smaller uncertainties, particularly for $\alpha_{1}$ and $\alpha_{2}$. The inclusion of detuning as an additional parameter in the response fit increases the standard error. This demonstrates the increased robustness of the backbone method against resonance-frequency drifts and standard characterization methods.
}
\end{table*}

The main advantage of the backbone method is the accurate determination of the nonlinear parameters while also acknowledging potential frequency drifts. To quantify this accuracy, we present the fit results as well as their relative error in Tab.~\ref{tab:fitcomparison}. Here, the values of the nonlinear coefficients extracted from frequency response measurements with and without an additional detuning parameter accounting for resonance frequency drift, as well as from the simultaneous two-parameter backbone fit, are given. The backbone method proves to be the more accurate approach, as it yields significantly smaller standard error, especially for the estimation of $\alpha_{1}$ and $\alpha_{2}$. For $\alpha_{2}$, the fit of the frequency response with drift compensations reaches a relative error of $118\%$.

For $\alpha_{3}$, the relative error obtained from the backbone fit ($4.3\%$) is comparable to that of the one-parameter response fit ($3.5\%$). This can be understood from the fact that the contribution of $\alpha_{3}$ to the frequency shift scales as $A^6$. Consequently, the information relevant for determining $\alpha_{3}$ is concentrated in the highest-amplitude part of the ringdown, such that only the earliest points after switch-off contribute significantly to its estimation. In contrast, the lower-order coefficients $\alpha_{1}$ and $\alpha_{2}$ influence a much larger portion of the decay and therefore benefit more strongly from the backbone analysis.

\section{Estimation of the Ringdown Start Time}
\label{app:ringstart}

\begin{figure}[ht!]
    \includegraphics[]{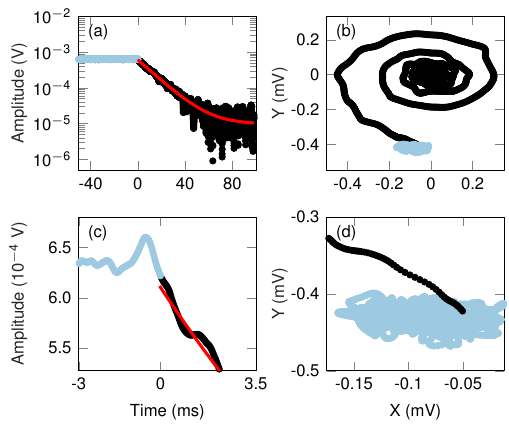}
    \caption{(a) Ringdown measurement reproduced from Fig.~\ref{fig:-35_-30}(a) of the main text. (b) Phase space representation of the same ringdown measurement reproduced from Fig.~\ref{fig:-35_-30}(b) of the main text. The turquoise dots represent the driven state, whereas the black dots denote the decaying response.
    (c) Close-up view of the amplitude around the start of the ringdown at the same drive power. The amplitude shows small amplitude fluctuations around the mean value, thus making the time estimate for the start of the ringdown ambiguous. (d) Close-up view of the phase space representation. Here, the transition between the cloud of turquoise dots and the black dots representing the ringdown is clear. Therefore, the phase-space representation can be used to determine the ringdown start time with very high accuracy. }
    \label{fig:-35close}
\end{figure}

As described in the main text, in addition to being key to our method for obtaining the backbone, recording the in-phase and quadrature signals also enables us to accurately estimate the start time of the ringdown. To fully visualize this, the amplitude and phase space representation of a ringdown measured with a drive power of \SI{-35}{\dBm}, also depicted in Fig.~\ref{fig:-35_-30}(a)-(b) of the main text, are reproduced in Fig.~\ref{fig:-35close}(a)-(b) along with a close-up view near the time where the drive is turned off in Fig.~\ref{fig:-35close}(c)-(d). As one can see from Fig.~\ref{fig:-35close}(c), the demodulated amplitude is not perfectly flat but exhibits ripples, thus making the visual estimation of the ringdown start time $t_0$ ambiguous. However, the phase space representation in Fig.~\ref{fig:-35close}(d) resolves this ambiguity as the distinction between the cloud of dots representing the driven and the undriven decaying state is clear and can be used to determine $t_0$ with high accuracy.     

\clearpage

\bibliography{References}
\end{document}